\documentstyle[11pt,aaspp4]{article}
 
\slugcomment{ApJ, in press}

\lefthead{Leggett, Allard, Hauschildt}
\righthead{Infrared Colors}
 
\begin{document}
 
\title{Infrared Colors at the Stellar/Substellar Boundary}

\author{S.K. Leggett}
\affil{Joint Astronomy Centre, University Park, Hilo, HI 96720
\nl skl@jach.hawaii.edu}
 
\author{F. Allard} 
\affil{CRAL, Ecole Normale Superieure, 46 Allee d'Italie, Lyon, 69364 France
\nl  fallard@cral.ens-lyon.fr  
\altaffilmark{1}}

\and
 
\author{P.H. Hauschildt}
\affil{Department of Physics and Astronomy \& Center for Simulational Physics, 
\nl University of Georgia, Athens, GA 30602-2451
\nl yeti@hobbes.physast.uga.edu}

\altaffiltext{1}{On leave from Wichita State University, Wichita, KS 67260}

\begin{center}
{\em full text with tables and figures available at\\
\tt ftp://ftp.jach.hawaii.edu/pub/ukirt/skl/dM\_preprint/}
\end{center}

\begin{abstract}
We present new infrared JHK photometry for 61 halo and disk stars 
around the stellar/substellar boundary.
We also present new L$^{\prime}$ photometry for 21 of 
these stars and for 40 low--mass stars taken from the Leggett 1992 
photometry compilation.  These data are combined with available
optical photometry and astrometric data to produce color--color 
and absolute magnitude--color diagrams --- the current sample extends 
the similar work presented in the 1992 paper into more metal--poor and
lower mass regimes.  The disk and halo sequences are compared to
the predictions of the latest model atmospheres
and structural models.

We find good agreement between observation and theory except for
known problems in the V and H passbands probably due to incomplete
molecular data for TiO, metal hydrides and H$_2$O.
The metal--poor M subdwarfs are well matched by the models 
as oxide opacity sources are less important in this case.
The known extreme M subdwarfs have
metallicities about one--hundredth solar, and the coolest subdwarfs
have T$_{eff}\sim 3000$~K with masses $\sim $0.09M/M$_{\odot}$. 
The grainless models are not able to reproduce the flux distributions of 
disk objects with T$_{eff} <$ 2500~K, however a preliminary version of the 
NextGen--Dusty models which includes homogeneous formation and extinction
by dust grains  {\it is} able to match the 
colors of these very cool objects.  The least luminous objects in this
sample are GD165B,  three DENIS objects ---
DBD0205, DBD1058 and DBD1228  --- and Kelu-1. These have T$_{eff}\sim$ 2000~K 
and are at or below the stellar limit with masses 
$\leq$0.075M/M$_{\odot}$.  Photometry alone cannot constrain these parameters
further as the age is unknown, but published lithium 
detections for two of these objects (Kelu-1 and DBD1228)  imply that they
are young (aged about 1 Gyr) and substellar (mass $\leq$0.06M/M$_{\odot}$).

\end{abstract}
 
\section{Introduction}

Low--mass stars are the most numerous type of object in the solar
neighbourhood.  Studies of the parameters of these long--lived objects --- 
surface gravity, temperature, metallicity, luminosity, radius
and mass --- are important for our understanding of the
Galactic mass density and its chemical evolution.  In Leggett (1992, hereafter 
L92) photometric and kinematic data were compiled for about 300 red dwarfs
and sequences were identified in color--color diagrams that were 
representative of the halo and the disk populations.  These allowed
empirical determinations to be made of metallicity and luminosity, 
based on observational colors.

Until very recently the observational data for low--mass stars could not be 
well reproduced by
synthetic spectra or photometry.  The cool, high--pressure 
atmospheres are difficult to model, due in particular to complex opacity 
sources: strong molecular bands and, for the halo stars and very low--mass 
objects, pressure--induced molecular hydrogen opacity (see e.g.
\cite{bor97}). The situation is now much
improved as demonstrated by for example the ``NextGen'' models of 
\cite{all98b} (see also \cite{all97}).
The observational side of the study of low--mass stars  has also changed
remarkably since 1992, with a large increase in the known number of
low--mass stars, brown dwarfs, and even giant planets.  This growth is 
due to the continuing dedicated effort by many groups, and to improvements
in sensitivity and accuracy of the various techniques employed.

In this work we present new infrared photometry for a sample of halo and disk 
stars approaching and even below the stellar/sub--stellar boundary.   This
work extends 
L92 to more metal--poor and lower mass regimes.  Also the data are taken 
with a well known photometric system, removing the need for uncertain color 
transformations. The disk and halo sequences are compared to synthetic colors
from Allard \& Hauschildt's NextGen model atmospheres (\cite{all98b}) and to
isochrones from Baraffe \& Chabrier's structural models 
(\cite{bar97,cha97}) combined with the NextGen atmospheres.  We also compare
the data to colors from a preliminary version of NextGen which includes grain 
condensation (\cite{all98c}).

In \S 2 we describe the sample for which new photometry is presented in \S
3.  \S 4 describes the model atmospheres and also describes the calculation of 
the synthetic photometry.  In \S 5 the model predictions are compared to the 
observations, and metallicity, mass and effective temperature are estimated 
for the sample.  Our conclusions are given in \S 6.  The Appendix gives
information on the CIT, IRTF and UKIRT photometric systems.

\section{The Sample}

JHK photometry was obtained for a selection of the more recently identified
very low--mass stars (or brown dwarfs) of the halo and disk.  The halo
stars were selected
from studies of known high proper--motion stars by \cite{g97,gr97}
and \cite{m92}.  The disk stars were selected from publications of new
discoveries by various color or proper--motion surveys --- 
\cite{d97} (the DENIS survey
DBD objects); \cite{i91} (the BRI objects); \cite{k96} (the 2MASS survey);  
\cite{rla97} (Kelu-1); \cite{s91} (PC0025); \cite{t93} (the TVLM objects) ---
or from follow--up studies of known high proper--motion stars --- 
\cite{k97}; \cite{m92} --- or from studies of low--mass companions to
nearby stars --- \cite{bz88} (GD165B).  The sample also includes four red
dwarf companions to cool white dwarfs observed in the course of the white 
dwarf work by \cite{ber97} (LP356-87, LHS~229, G111-72 and G59-39). 
The target stars are listed in Table 1. We give LHS or LP number 
(\cite{luy79}),
and/or Gliese or Gliese/Jahreiss number (\cite{gj91}), and/or Giclas
number (\cite{g71}), for each star.  We also give discovery names for the
objects taken from the sources cited above, where we have abbreviated 
``TVLM'' to ``T''.  An abbreviated RA/Dec is also given to aid
identification.  

The spectral types in Table 1 are taken from various sources and there may
be discrepancies or errors at the level of one sub--class.  For the halo
stars the classifications primarily are from \cite{g97} and \cite{gr97}, but 
for LHS~2045, LHS~2245, LHS~3168 and LHS~3390 they are based on our own
optical spectra.  For the disk stars the classifications are taken 
from \cite{g97,k95} and \cite{k97}, and we have classified the new 
objects Kelu-1 and the DENIS objects by this scheme, based on published
optical spectra.

The sources for the astrometric data --- trigonometric
parallax and radial velocity --- are given in Table 1.  We have calculated
the UVW space motions for each object that has both parallax and radial 
velocity measurements, and classified these objects into kinematic
populations of
young disk (YD), young/old disk (Y/O), old disk (OD), old disk/halo (O/H) 
and halo (H).  We have followed the velocity specifications described in L92 for
these classifications.

\section{Observations}

We obtained JHK data for the stars in Table 1 using infrared cameras
on two telescopes, both situated on Mauna Kea in Hawaii.  These cameras
were NSFCAM on the NASA Infrared Telescope Facility (\cite{shu94}) and
IRCAM on the UK Infrared Telescope (\cite{pux94}) both of which use InSb 
detectors. The JHKL$^{\prime}$ filter sets for
the instruments are effectively identical except for the H filter.
The only color transformation required is at H and the correction onto the 
UKIRT system for those objects observed only with NSFCAM is 
$\leq$7\%.  We give the 
NSFCAM--IRCAM (IRTF--UKIRT) H--band color transformation in the 
Appendix.  UKIRT Faint Standards were used to calibrate
the JHK photometry and  UKIRT Bright Standards were used for L$^{\prime}$
(\cite{cas92}, see also URL 
http://www.jach.hawaii.edu/UKIRT/astronomy/standards.html); 
the Appendix gives some more information on the UKIRT system.

Table 2 lists new JHK data for the 61 stars in our sample where 
we have converted the NSFCAM H--band data to the UKIRT system.  
We include data for LHS~377 although these data were obtained in 1994 and 
published in \cite{l96}; we list it here again for reference, as it was not 
included in L92. The other data 
were obtained on various telescope runs over the period 1995 to 1998; the 
camera used and the number of observations are given in Table 2. 
The technique used was the standard ``jitter'' technique, also known as 
``dithering''.  This technique involves making multiple exposures,
typically 5, with the telescope offset between each exposure such that 
the star is placed on different regions of the array.  Based on the
reproducibility of the photometry between jitters the errors at JHK are
3\% for stars with K$<$14 and 5\% for the fainter stars.

Table 2 also gives new L$^{\prime}$ photometry for 21 stars,
and more new L$^{\prime}$ data are given in Table 3 for 40 stars in the L92
sample.  These data were also taken with either NSFCAM or IRCAM,
for which the filter and photometric systems are identical.  The data
were obtained during the period 1995 to 1998, except for LHS~377 and 
LHS~248 for which  L$^{\prime}$ photometry was obtained through the UKIRT 
Service Observing Program in 1994.  The L$^{\prime}$ data for these two 
stars were published in
\cite{l96} but we include the data here for reference.  Errors are given
for L$^{\prime}$  for each star in Tables 2 and 3.

Table 4 gives VIJHKL$^{\prime}$ colors for the sample of 61 stars
specified in Table 1.
The optical data are taken from various publications as listed in the
table, and are on the Cousins system (the data from \cite{w87} and
\cite{e79}
were converted to the Cousins system using the relationships summarised
in L92). We also give the distance modulus $M-m$ determined from the
parallax given in Table 1.

\section{Models}

For comparison with the observed photometry we show the results of two sets of
models, the grainless NextGen models (\cite{all98b})
and the NextGen--Dusty models 
(\cite{all98c}).  These models have been computed with
the stellar atmosphere code PHOENIX 
(version 9.1) in its static, plane--parallel, LTE mode (see e.g.
\cite{hau97}). 
The metallicities used are all simple scaled solar 
metallicity values, with, for example,  no enhancement of $\alpha$ elements.

\subsection{NextGen}

The NextGen models are described in  \cite{all97} and  \cite{all98b}.  
The equation of state for these models is an enlarged and enhanced version of 
that presented in \cite{all95}.  Also, new molecular line data have been 
incorporated.  These include the lines from the HITRAN92 (\cite{hitran92})
database, line lists for YO (\cite{yolines}) and ZrO (\cite{zrolines}),
the \cite{TiOJorg} list of TiO lines (as well as other lists from his
SCAN database), and the available list of water lines of Miller \& Tennyson 
as described in \cite{h2olet}.  

Both atomic and molecular lines are treated with a direct opacity
sampling method.  We do {\em not} use pre--computed opacity sampling
tables, but instead dynamically select the relevant LTE background lines
from master line lists at the beginning of each iteration for every
model and sum the contribution of every line within a search window to
compute the total line opacity at {\em arbitrary} wavelength points.
In the calculations we present in this paper, we  have included a
constant statistical velocity field, $\xi =$ 2~km/s, which is treated as
a microturbulence.  The choice of lines is dictated by whether they are
stronger than a threshold $\Gamma\equiv \chi_l/\kappa_c=10^{-4}$, where
$\chi_l$ is the extinction coefficient of the line at the line center
and $\kappa_c$ is the local bound--free absorption coefficient.  This typically
leads to about $20\cdot 10^{6}$ lines which are selected from the master line
lists (with 47 million atomic and up to 340 million molecular lines).
The profiles of these lines are assumed to be depth--dependent Voigt profiles 
(or Doppler for very weak lines).  Details of the computation of the
damping constants and the line profiles are given in \cite{vb10pap}.

\subsection{NextGen--Dusty}

The NextGen--Dusty models have been introduced in \cite{all97} and a detailed
publication is in progress  (\cite{all98c}).  These models include both the 
formation of and the extinction by dust grains (absorption and
scattering) --- they are present both in the equation
of state and as an opacity source.  The chemical equilibrium has been augmented
to include, besides a complete gas phase composition of over 600 species, the 
formation of over a thousand crystal and liquid condensates. These so-called 
dust grains are assumed to be in equilibrium with the surrounding gas phase 
medium, and their abundance is computed using the Gibbs energy of formation
compiled in the JANAF thermochemical tables (\cite{janaf}).  

These models assume an
interstellar grain size distribution and reconstruct the grain opacity
coefficients as a function of wavelength using polarizability spectra
from the literature and the Continuous Distribution of Ellipsoids
theory of \cite{boh83}. The opacity  profiles (both 
absorption and scattering) of twelve types of grain have been 
included --- Fe, Ni, Cu, corundum
(Al$_2$O$_3$), MgAl$_2$O$_3$, enstatite (MgSiO$_3$), forsterite
(Mg$_2$SiO$_4$), and the silicates CaSiO$_3$, Ca$_2$SiO$_4$,
CaMgSi$_2$O$_6$, Ca$_2$Al$_2$SiO$_7$, and Ca$_2$MgSi$_2$O$_7$.  These are
the most abundant condensates in late--type M
dwarfs.  The opacity of three of these grain species (Fe, forsterite,
and Ca$_2$MgSi$_2$O$_7$) had however to be omitted to reproduce the
infrared spectral distribution of late--type red dwarfs (see e.g. \cite{rla97}),
suggesting that these grains have settled gravitationally below the 
photosphere, or have failed the nucleation process.  Further discussion
is beyond the scope of this paper, but will be presented in  \cite{all98c}. 
The Ruiz et al. work showed that inclusion of these particular species (and only
these species) leads to an excessive amount of infrared flux compared to the
observed energy distribution of Kelu-1 (the effect of 
grains is to heat the upper layers
of the atmosphere and to disassociate H$_2$O).  This work shows that Kelu-1 
does not have unusual colors and these species must be omitted for all 
late--type red dwarfs.

A further improvement  on
the NextGen grid is the use of the latest water vapor line
list by \cite{ps97} which includes over 300 million
transitions, and is more complete for the higher energy transitions than
the Miller \& Tennyson line list.  The Partridge \& Schwenke list has been
used for the models with T$_{eff}<$2000~K and has the effect of increasing
the depth of the water bands, compared to the less complete Miller \& Tennyson 
line list (but this effect is offset by that of the inclusion of grains). 

The NextGen--Dusty model results presented in this work are preliminary, but 
they do demonstrate the likely effect of the inclusion of grains in the 
atmospheres of very low mass stars and brown dwarfs.  The diagrams in the 
following sections show that these preliminary dusty models do a very good job
of reproducing the observed colors of low--mass objects, and that the grainless
models fail for T$_{eff}<$2500~K.  As the dusty models are preliminary, we adopt
single values of surface gravity and metallicity given by log~$g=5.0$ and 
[m/H]$=$0.  Grains will also be important for lower metallicities; initial 
calculations show that for [m/H]$=-0.5$ condensation affects the
spectral energy distribution for   T$_{eff}<$2300~K and for  [m/H]$=-1.0$
it is important for T$_{eff}<$2000~K.  However there are no metal--poor
stars in our sample at such low temperatures and we defer presentation
of these results to \cite{all98c}.

\subsection{Synthetic Photometry}

The JHKL$^{\prime}$ colors on the UKIRT system were calculated  by integrating 
the synthetic spectra over the UKIRT bandpasses (see the Appendix for 
more information on the UKIRT system).  The VI color computations used the 
Cousins V and I bandpasses as given by \cite{bes90}.  Terrestrial atmospheric 
absorption was included in all filter bandpasses.  The magnitudes were 
calibrated by dividing the synthetic spectra by the observed spectrum of Vega 
(\cite{hay85,mou85}), which was adopted to define zero magnitudes at all 
wavelengths. 

\subsection{Structural Models and Absolute Magnitudes}

Structural and evolutionary models for very low mass stars have recently 
been calculated --- by \cite{bar97} and \cite{cha97} --- which use updated
physics
for these dense and cool objects together with the NextGen (grainless)
atmospheres.  These isochrones provide a theoretical relationship between
effective temperature and surface gravity, at a given metallicity, mass and 
age.  Synthetic color grids have been calculated from the NextGen synthetic 
spectra
for a range of effective temperature, metallicity and surface gravity, and 
these grids were interpolated to produce colors and absolute magnitudes
at the appropriate effective temperature and gravity for a mass sequence
at a given metallicity and age.  Synthetic color grids have also been 
calculated from the NextGen--Dusty models but these have not yet been
combined with the structural models, and  for these models we adopt 
single values of surface gravity and metallicity given by log~$g=5.0$
and [m/H]$=$0.  Where absolute magnitudes are presented for these dusty
models in the diagrams below, these magnitudes have been derived by scaling 
the calculated stellar surface magnitudes to match the NextGen (grainless)
absolute magnitudes at 2900K.  At this temperature condensation is not 
important (but below this temperature it is).

We have investigated whether gravity can be reliably determined from the 
observed photometry, as this could provide an independent check on the 
theoretical T$_{eff}$--log~$g$ relationship used in the evolutionary
models, or a check on the age of the stars in the 
sample.   For the disk stars the structural models imply that
log~$g$ ranges from 4.7 at the hot/massive end to 5.4
at the cool/low--mass end, for an age of 10~Gyr.  For a younger age of 1~Gyr,
and solar metallicity, the low--mass value is slightly lower --- 
log~$g \sim$ 5.3. For the halo stars the range is 5.1 $\leq$log$~g \leq$
5.5.  We found that for a reasonably large change in gravity 
of 0.5~dex the synthetic colors differ by an amount less than the
observational error, and so we cannot constrain the theoretical
T$_{eff}$--log~$g$ relationship or the age of the stars.

\section{Comparison of Data and Models}

\subsection{Color--Color Diagrams}

In all the diagrams in this subsection we show colors generated by
interpolating the NextGen model color grid onto the T$_{eff}$--log~$g$
values generated by model isochrones, as described above.  We use a fixed 
age of 10~Gyr as colors alone are insensitive to varying age 
(while keeping metallicity constant), except for the lowest mass objects
at very young ages (results for an age of 1~Gyr are shown in the absolute
magnitude--color diagrams presented in the following section).  The 
NextGen--Dusty colors  have been generated for a single value of log~$g=5.0$.  
This simplification to a single value for surface gravity is valid
as the colors are not a strong function of gravity (for example, at 
T$_{eff}=$1900~K changing log~$g$ by 0.5~dex only changes the various colors
plotted in the following diagrams by $\leq$0.10~mag).

In all the plots the filled symbols are
new data from the current work and the open symbols are from L92. The L92
data have been converted to the UKIRT JHK photometric system
using transformations given by \cite{cas92} (see Appendix). 
Symbol shapes illustrate kinematic populations:
squares are halo, triangles are disk 
(including all young disk to old disk/halo) and circles are unknown
kinematic population.

Figure 1 shows the V$-$I:I$-$K diagram with NextGen model sequences for
metallicities [m/H]$=-2, -1$ and 0 and the NextGen--Dusty colors for  
log~$g=5.0$ and [m/H]$=0$.  Lines of constant effective temperature
are superposed. A typical observational error bar is shown, as well as the 
larger error bars for Kelu-1.  
It can be seen that while the metal--poor halo sequence
appears to be well reproduced (the extreme subdwarf LHS~205a is labelled), 
there is a problem for the disk sequence.
Other color--color diagrams show that the problem is in the V band, and in
this spectral region there are known inadequacies in the metal hydride and
titanium oxide spectroscopic data (\cite{alv97}).  It is not known whether
these molecular data are the source of the error, whether there is some
other form of missing opacity, or whether the problem is structural, such
as the treatment of convection.  Although problems in one spectral
region imply that an overall redistribution of the stellar flux is needed, 
very little flux is emitted at V for the red dwarfs --- tests on the model
atmospheres indicate that even a 0.5 mag additional opacity in the V bandpass
would not cause significant flux redistribution to JHK wavelengths 
(\cite{bar98}).

Figure 2 shows I$-$J:J$-$K which is a more 
complete representation of our data sample (only eight stars do not have I data 
available compared to twenty lacking V, some of which are the reddest 
objects).  Again a typical error bar is shown as well as the larger
errors for Kelu-1 and the DENIS objects.  The  cool extreme subdwarfs
are labelled as well as the coolest disk objects.  Note that the famous 
non--controversial brown dwarf Gl~229B (\cite{nak95}) is not shown as it has 
very different colors 
due to strong methane absorption in the H and K bands (I$\sim 20.0$, 
J$\sim$H$\sim$K$\sim$14.3, \cite{mat96}).  Spectral types for the disk
stars are given on the right--hand axis of Figure 5.  These are based on
the relationship between I$-$J and spectral class demonstrated in L92, and 
they agree with the individual classifications listed 
in Table 1.   The recently discovered extremely red dwarfs have raised the 
issue of classification beyond M9, as discussed by \cite{k97a}. 

Although it is known that the Miller \& Tennyson line list for water, the
most important near--infrared opacity source, is incomplete
(\cite{all97}), 
the agreement  between theory and observation is good for these colors ---
which  cover the peak in the stellar energy distribution.  For the coolest
objects the agreement becomes remarkably good when the effect of grain
formation and extinction is taken into account.  At the blue end of the 
sequence there is a discrepancy at  the 10\% level  which is
not seen if the L92 data is matched in CIT--system space via other filter
profiles (\cite{all98a}).  Either there is an error at this level in the
models, which is cancelled by errors in the CIT color transformation, or
there are problems with the UKIRT defined system.  One potential source of 
error is the UKIRT J filter which has
its red transmission edge determined by telluric water vapor absorption. 
However computations using a realistic range of telluric water vapor show
that photometric deviations of only about 3\% are likely.
This problem should be resolved with 
improvements to infrared photometric systems (such as the filter set now 
being purchased by a Mauna Kea Observatory consortium which is better 
matched to the atmospheric windows) and further refinements to the models.

Figure 3 is the J$-$H:H$-$K diagram in the UKIRT system.  We also indicate
the CIT system values on opposite axes for comparison to other published 
work.  We have labelled the bluest (in J$-$H) metal--poor M dwarfs 
as well as the reddest disk objects.  This diagram is still very difficult for
the models to reproduce, at least for disk stars.  The problem here is
predominantly the H band where the overall opacity is lower, and hence
photons are originating in a regime of higher temperature and pressure
--- where the Miller \& Tennyson H$_2$O opacity profile is poorly defined.   
Nevertheless the 
models present a clear improvement in this diagram compared to previous model 
generations. Metal--poor models match the behavior of the most extreme 
subdwarfs,  with colors becoming bluer as the effective temperature decreases
(due to pressure--induced H$_2$ absorption in the K bandpass), 
while metal--rich models and stars are confined to red J-H values.
Dust formation accounts for the extreme
redness of the least massive metal--rich stars. 

Figure 4 shows the longer wavelength  J$-$K:K$-$L$^{\prime}$. The outliers
LHS~375 --- an extreme subdwarf --- and Kelu-1--- an extremely cool disk 
object --- are labelled.  L$^{\prime}$ photometry is only available for the 
brighter members of this and the L92 sample.  The agreement
is good especially when 
grain opacities are included for the coolest objects.

\subsection{HR Diagrams}

In these diagrams the symbols are the same as in the earlier figures. In 
addition,
large crosses show the location of the end of the main sequence, based on
\cite{bar97} and \cite{cha97} models.  The coolest subdwarfs and disk
objects are labelled.
   
The parallaxes for 
the L92 data (open symbols) have been updated using the final version
of the Yale Catalogue (\cite{van94}) resulting in small changes to the
observed absolute magnitudes for this sample.  Vertical error bars 
are shown for stars with parallax errors greater 
than 20\%. Note that these diagrams show that 
the parallax for the distant pair TVLM514-42404A/B has been
overestimated.  Only the 50 of the 61 stars
in the current sample with parallax measurements available are plotted, but
not all of these appear on all the diagrams due to incomplete color coverage ---
the redder colors are more complete.

In absolute magnitude--color plots such as these the metal--poor stars 
form what appears to be a ``sub--luminous'' sequence
sitting below the main sequence --- hence the name ``subdwarfs''.
In fact, as is well known, the 
metal--poor halo stars are hotter and bluer than more metal--rich
stars of the same mass, due to decreased optical and increased infrared 
opacity. The exact shift with 
metallicity is a function of the color plotted and the mass of the star.
Although there is not a 
one--to--one correspondence between space motion and metallicity, the halo 
kinematic sample clearly defines a ``sub--luminous'' sequence in these figures.  

Figure 5 shows M$_V$:V$-$I with 10~Gyr isochrones superimposed for
metallicities [m/H]$=-2, -1$ and 0.  A dotted line also shows the solar
metallicity 1~Gyr isochrone, however this is indistinguishable from its
older counterpart as no structural differences are expected except for
masses lower than 0.09M/M$_{\odot}$ and M$_V >$17, at which point younger
objects of the same mass will be brighter.  Lines of constant mass are
shown.  The model discrepancy in the V band for
the more metal--rich stars is again apparent.

Figure 6 shows the better reproduced M$_I$:I$-$J, and Figure 7 shows
M$_J$:J$-$K where the right axis shows empirical masses as a function of 
absolute J magnitude, based on the relationship determined by \cite{hm93} for
low--mass binary systems.  These masses are valid for the disk stars only.
The dotted line is the model sequence for an age of 1~Gyr with the 
hydrogen burning limit corresponding to brighter magnitudes compared to
the older sequence.  Despite the $\sim$0.1~mag offset seen in J$-$K, as
discussed in \S 5.1, the 
match between observation and theory is reasonable, especially when grains are 
included for the cool disk objects.  The theoretical masses agree well
with those derived observationally by Henry \& McCarthy. 

\subsection{Results}

Table 5 gives estimates of metallicity, mass and effective temperature for
the stars in our sample, based on the model comparisons shown in Figures 1---7.
These classifications were performed by eye, taking into account observational 
error and apparent systematic discrepancies between theory and observation at 
V$-$I and J$-$K, to produce a consistent estimate of metallicity and 
effective temperature.  We have grouped
the stars first into broad metallicity classes of [m/H]$\sim -2, -1,$ and
0, and listed them in order of decreasing effective temperature within
these groups.  We indicate values of T$_{eff}$ every 500~K, with the
corresponding mass as implied by the structural models of
\cite{bar97} and \cite{cha97}. 

These values of metallicity, mass and effective temperature
can only be estimates as spectroscopy is required for a more accurate
determination of stellar parameters, and for the lowest mass objects age is
also important (and unknown).  Also, it is likely that the absolute 
values determined by model fitting will change in the near future as the
models are upgraded.  Nevertheless the overall 
trends indicated in Table 5 will be correct.  

Spectroscopic lithium detections have been recently published for three of the 
coolest objects in this sample --- Kelu-1 (\cite{rla97,bas97}), DBD1228 
(\cite{bas97,t97}), and LP944-20 (\cite{t98a}).  The detection of lithium
combined with determinations of T$_{eff}\sim 2000$~K for these three objects 
implies that their ages are about 1~Gyr and necessarily constrains their
masses to be $\leq$0.06M/M$_{\odot}$, i.e. substellar,  based on current 
models by \cite{cha97}.  This mass constraint arises from the fact that
lithium depletion timescales are short, so for a star to show lithium 
it has to be either well below the stellar mass limit or it has
to be extremely young (aged about 0.1~Gyr). However if it is extremely
young {\it and} massive enough to be stellar then  \cite{cha97} calculate
that T$_{eff}>2800$~K, i.e. much hotter than these
objects.  The age constraint arises from the fact that a substellar object
would have cooled below 2000~K if the age were greater than about 1~Gyr.  
Finally, the fact that LP944-20 and Kelu-1 have
partially depleted their lithium implies they are not significantly younger 
than 1~Gyr.  

\cite{bas97} also looked for lithium in DBD1058 but failed to detect
it.  The depletion of lithium and the fact that  T$_{eff}\sim2000$~K, implies a
mass for this object between 0.06M/M$_{\odot}$, if it is aged about 1~Gyr, and
0.075M/M$_{\odot}$ (the stellar limit), if it is aged several Gyr.  

Amongst the metal--poor stars LHS~453 and LP~251-35 are notable as
they are the most metal--poor stars in our sample with  [m/H]$<-2$, based
on the comparisons with the models presented 
in this work.  These stars are known extreme subdwarfs.  Another metal--poor 
star of particular interest is LHS~254, which is very red.  Our
photometry implies that it may be as metal--poor as LHS~377 with [m/H]$=-1$ 
and hence
could be very close to the hydrogen--burning limit for that metallicity.
Follow up observations are required, including measurement of its 
trigonometric parallax (such work is in progress at the US Naval Observatory 
Flagstaff Station).

\section{Conclusions}

The last two years have seen remarkable advances in the study of very
low--mass stars and brown dwarfs.  These advances have been on both the
observational front, where we now have well defined sequences for both the
halo and disk extending close to or beyond the brown dwarf boundary,  and
on the theoretical front, with much more complex and realistic models for
both the stellar atmosphere and structure.

We have presented new infrared data for a sample of the low mass
metal--poor stars and the very cool disk objects.  These data are all on
the same photometric system and allow the halo and disk sequences to
be identified down to the hydrogen burning boundary.  Although halo
stars at the stellar limit have not yet been identified, the current
sample includes halo red dwarfs with masses down to 
$\sim$0.09M/M$_{\odot}$ and T$_{eff} \leq$ 3000~K.  The disk sample
includes lithium--confirmed field brown dwarfs --- Kelu-1
(\cite{rla97,bas97}), DBD1228 (\cite{bas97,t97}) and LP~944-20
(\cite{t98a}).

Comparison to theoretical models gives good agreement in
color--color diagrams, especially when grain 
condensation is included.  Remaining problem areas are being worked on
--- improvements to the line lists for molecular species such as TiO,
hydrides and H$_2$O, and inclusion of the gravitational settling 
of grains --- and we can look
forward to improvements in the match to the observational data in the near
future.  A more detailed spectroscopic comparison between observation and
theory will be presented by this group in a subsequent paper.

We can also no doubt look forward to a continuing increase in the number
of known low mass objects as observational efforts continue, including
the large--scale sky surveys by the DENIS group (\cite{d97}) and 2MASS 
(\cite{k96}), which are now in full production mode.

\acknowledgments
We are very grateful to the staff at IRTF and UKIRT for their 
assistance in obtaining the data presented in this paper.  UKIRT, the United 
Kingdom 
Infrared Telescope, is operated by the Joint Astronomy Centre on behalf
of the U.K. Particle Physics and Astronomy Research Council.  The IRTF, 
the NASA Infrared Telescope Facility, is operated by the University of Hawaii 
under contract to NASA.  The UKIRT Service program also contributed to
this work.  Isabelle Baraffe kindly interpolated the atmospheric models on
to the interior models to determine mass and radius at various ages, and
Alastair Glasse calculated the transmission of the Mauna Kea atmosphere as
a function of precipitable water vapor.  We are grateful to Chris Tinney and 
Neill Reid for early access to their radial velocity data, and to Allyn Smith 
and Terry Oswalt for permission to use unpublished optical data. Tom Geballe 
read through a draft of the paper and gave useful comments. 
FA acknowledges support from  NASA LTSA NAG5-3435 and NASA EPSCoR grants to 
Wichita State University.  PHH acknowledges partial support by
NASA ATP grant NAG 5-3018 and LTSA grant NAG 5-3619 to the University
of Georgia.  Some of the calculations
presented in this paper were performed on the IBM SP2 of the UGA UCNS,
at the San Diego Supercomputer Center (SDSC), the Cornell Theory Center
(CTC), and at the National Center for Supercomputing Applications (NCSA),
with support from the National Science Foundation.
We thank all these institutions for a generous
allocation of computer time.

\clearpage

\appendix
\section{Appendix --- Infrared Color Transformations}

In L92 all JHKL data were given on the CIT photometric system.  At that
time the CIT system was commonly used, with many infrared telescopes
using the standard stars on this system published by \cite{eli82}.  Since 
1992, however, more sensitive infrared array cameras have become 
the instruments of choice and the commonly used set of standards is now
the UKIRT Faint Standards List published by \cite{cas92}
(see also URL 
http://www.jach.hawaii.edu/UKIRT/astronomy/standards.html).
For this reason, as well as to avoid introducing errors through uncertain
color transformations, we have left the new data presented in this work on 
the UKIRT system.  Many color transformations were summarised in L92 and
here we wish to update the CIT to UKIRT transformations.  The most recent
determination is that published by \cite{cas92} and the following should 
be used in preference to the relations given in L92:
$$ K_{CIT} = K_{UKIRT} - [0.018 \times (J - K)] $$
$$ (J - K)_{CIT} = 0.936 \times (J - K)_{UKIRT} $$
$$ (H - K)_{CIT} = 0.960 \times (H - K)_{UKIRT} $$
$$ (J - H)_{CIT} = 0.920 \times (J - H)_{UKIRT} $$
Note that the original UKIRT Faint Standards data, and the transformation to
CIT, were obtained using a single aperture photometer UKT9.  However 
comparisons between UKT9 and IRCAM show no significant systematic difference 
for stars with $J - K < 1$ (\cite{cas92}; Hawarden, private communication).

Copies of the UKIRT filter profiles can be retrieved by  following the 
``available filters'' link from URL
http://www.jach.hawaii.edu/UKIRT/instruments/ircam/ircam3.html.
During the course of this work we determined that the IRTF's NSFCAM filters
are effectively identical to UKIRT's IRCAM filters at J, K and L$^{\prime}$, 
but not at H.  Observations
of red dwarfs show that the transformation required to bring IRTF--H onto 
the UKIRT system is:
$$ (H - K)_{UKIRT} = 0.82 [\pm 0.02] \times (H - K)_{IRTF} $$
which is valid for $ 0 \leq (H - K)_{UKIRT} \leq 0.5 $.

\clearpage 

\figcaption[skl_dm_fig1.eps]{V$-$I:I$-$K diagram with model sequences for 
metallicities [m/H]$=-2, -1$ and 0.  Lines of constant effective temperature
are shown.  Filled symbols --- 
this work, open --- L92 on the UKIRT JHK system.
Symbol shapes are kinematic populations: squares --- halo, triangles --- 
disk, circles --- unknown. \label{fig1}}

\figcaption[skl_dm_fig2.eps]{I$-$J:J$-$K with model sequences using the same 
symbols as in Figure 1.  Spectral types based on I$-$J (L92) are shown.
\label{fig2}}

\figcaption[skl_dm_fig3.eps]{J$-$H:H$-$K diagram with model sequences
using the same symbols as in Figure 1.\label{fig3}}

\figcaption[skl_dm_fig4.eps]{J$-$K:K$-$L$^{\prime}$ with model sequences
using the same symbols as in Figure 1.\label{fig4}}

\figcaption[skl_dm_fig5.eps]{M$_V$:V$-$I with 10 Gyr isochrones 
superimposed for
metallicities [m/H]$=-2, -1$ and 0.  Lines of constant mass are
shown.  The symbols are as in Figure 1.  The dotted line is the
1 Gyr isochrone, and large crosses indicate the end of the 
hydrogen burning sequences.\label{fig5}}

\figcaption[skl_dm_fig6.eps]{M$_I$:I$-$J, with symbols and line types as in 
Figure 5.\label{fig6}}

\figcaption[skl_dm_fig7.eps]{M$_J$:J$-$K,  with symbols and line types as in 
Figure 5. Empirical 
masses for the disk stars from \cite{hm93} are shown.\label{fig7}}

\end{document}